\input{psfig}
\documentstyle[onecolumn]{mn}

\newif\ifAMStwofonts

\ifoldfss
  \ifCUPmtlplainloaded \else
    \NewTextAlphabet{textbfit} {cmbxti10} {}
    \NewTextAlphabet{textbfss} {cmssbx10} {}
    \NewMathAlphabet{mathbfit} {cmbxti10} {} 
    \NewMathAlphabet{mathbfss} {cmssbx10} {} 
  \fi
  \ifAMStwofonts
    \ifCUPmtlplainloaded \else
      \NewSymbolFont{upmath} {eurm10}
      \NewSymbolFont{AMSa} {msam10}
      \NewMathSymbol{\upi}     {0}{upmath}{19}
      \NewMathSymbol{\umu}     {0}{upmath}{16}
      \NewMathSymbol{\upartial}{0}{upmath}{40}
      \NewMathSymbol{\leqslant}{3}{AMSa}{36}
      \NewMathSymbol{\geqslant}{3}{AMSa}{3E}

      \let\leq=\leqslant \let\le=\leqslant
      \let\geq=\geqslant \let\ge=\geqslant
    \fi
  \fi
\fi 

\ifnfssone
  \newmathalphabet{\mathit}
  \addtoversion{normal}{\mathit}{cmr}{m}{it}
  \addtoversion{bold}{\mathit}{cmr}{bx}{it}
  \newmathalphabet{\mathbfit} 
  \addtoversion{normal}{\mathbfit}{cmr}{bx}{it}
  \addtoversion{bold}{\mathbfit}{cmr}{bx}{it}
  \newmathalphabet{\mathbfss} 
  \addtoversion{normal}{\mathbfss}{cmss}{bx}{n}
  \addtoversion{bold}{\mathbfss}{cmss}{bx}{n}
  \ifAMStwofonts
    \ifCUPmtlplainloaded \else
      \UseAMStwoboldmath
      \makeatletter
      \new@mathgroup\upmath@group
      \define@mathgroup\mv@normal\upmath@group{eur}{m}{n}
      \define@mathgroup\mv@bold\upmath@group{eur}{b}{n}
      \edef\UPM{\hexnumber\upmath@group}
      \new@mathgroup\amsa@group
      \define@mathgroup\mv@normal\amsa@group{msa}{m}{n}
      \define@mathgroup\mv@bold\amsa@group{msa}{m}{n}
      \edef\AMSa{\hexnumber\amsa@group}
      \makeatother
      \mathchardef\upi="0\UPM19
      \mathchardef\umu="0\UPM16
      \mathchardef\upartial="0\UPM40
      \mathchardef\leqslant="3\AMSa36
      \mathchardef\geqslant="3\AMSa3E

      \let\leq=\leqslant \let\le=\leqslant
      \let\geq=\geqslant \let\ge=\geqslant
    \fi
  \fi
\fi 

\ifnfsstwo
  \DeclareMathAlphabet{\mathbfit}{OT1}{cmr}{bx}{it}
  \SetMathAlphabet\mathbfit{bold}{OT1}{cmr}{bx}{it}
  \DeclareMathAlphabet{\mathbfss}{OT1}{cmss}{bx}{n}
  \SetMathAlphabet\mathbfss{bold}{OT1}{cmss}{bx}{n}
  \ifAMStwofonts
    \ifCUPmtlplainloaded \else
      \DeclareSymbolFont{UPM}{U}{eur}{m}{n}
      \SetSymbolFont{UPM}{bold}{U}{eur}{b}{n}
      \DeclareSymbolFont{AMSa}{U}{msa}{m}{n}
      \DeclareMathSymbol{\upi}{0}{UPM}{"19}
      \DeclareMathSymbol{\umu}{0}{UPM}{"16}
      \DeclareMathSymbol{\upartial}{0}{UPM}{"40}
      \DeclareMathSymbol{\leqslant}{3}{AMSa}{"36}
      \DeclareMathSymbol{\geqslant}{3}{AMSa}{"3E}

      \let\leq=\leqslant \let\le=\leqslant
      \let\geq=\geqslant \let\ge=\geqslant
    \fi
  \fi
\fi 

\ifCUPmtlplainloaded \else
  \ifAMStwofonts \else 
    \def\upi{\pi}
    \def\umu{\mu}
    \def\upartial{\partial}
  \fi
\fi

\title{The pair-wise velocity dispersion of galaxies: effects
of non radial motions}
\author[A. Del Popolo]
       {A. Del Popolo \\
Dipartimento di Matematica, Universit\`{a} Statale di Bergamo,
Piazza Rosate, 2 - I 24129 Bergamo, ITALY\\
Feza G\"ursey Institute, P.O. Box 6 \c Cengelk\"oy, Istanbul, Turkey\\
Bo$\breve{g}azi$\c{c}i University, 80185 Bebek, Istanbul, Turkey
}
\date{}

\pagerange{\pageref{firstpage}--\pageref{lastpage}}
\pubyear{}

\begin{document}

\maketitle

\label{firstpage}

\begin{abstract}

I discuss the effect of non-radial motions on the
small-scale pairwise
peculiar velocity dispersions of galaxies
(PVD) in a CDM model. I calculate
the PVD for the SCDM
model by means of the refined
cosmic virial theorem (CVT) (Suto \& Jing 1997 (hereafter SJ97)) and taking account of
non-radial motions by means of Del Popolo \& Gambera (1998)
(hereafter DG98) model. I
compare the results of the present model with the data from Davis \&
Peebles (1983),
the IRAS value at $1 h^{-1} {\rm Mpc}$ of
Fisher et al. (1993) and Marzke et al. (1995). I show that while the SCDM
model disagrees with the observed values, as pointed out by several authors
(Peebles 1976, 1980; Davis \& Peebles 1983 (hereafter DP83); Mo et. al 1993;
Fisher et al. 1994b; SJ97; Jing et al. 1998 (hereafter J98)), taking account of
non-radial motions produce smaller values for the PVD.
At $r \leq 1 h^{-1} {\rm Mpc}$ the result is in agreement with
Bartlett \& Blanchard (1996) (hereafter BB96).
At the light
of this last paper, the result may be also read as
a strong dependence of the CVT
prediction on the model chosen to describe the mass distribution
around galaxies, suggesting 
that the CVT cannot be taken as a direct evidence of a low density
universe.
Similarly
to what shown in Del Popolo \& Gambera (1999, 2000) (hereafter DG99, DG00),
Del Popolo et al. (1999) (hereafter D99), the agreement of our model
to the observational data is due to a scale dependent bias induced by
the presence of non-radial motions. Since the assumptions on which
CVT is based have been questioned by several authors
(BB96;
SJ97),
I also calculated the PVD using
the redshift distortion in the redshift-space correlation function,
$\xi_{\rm z}(r_{\rm p},\pi)$, and I compared it with the PVD measured from
the Las Campanas Redshift Survey by J98. The result
confirms that non-radial motions influence the PVD making them better
agree with observed data.
\end{abstract}

\begin{keywords}
cosmology: theory - large scale structure of Universe - galaxies: formation
\end{keywords}

\section{Introduction}

The pairwise velocity dispersion of galaxies (PVD)
is an important quantity that gives information on the structure and
clustering of the universe. Peculiar velocities,
originating from the action of gravitational fields,
are a probe of the gravitational potentials produced by luminous
and dark matter.

Several approaches have been developed to determine the PVD.
In this paper, I'll be concerned with two
of them, the first one
based on the redshift-space
distortions and the second one on the Cosmic Virial Theorem (hereafter CVT). \\
\indent The first method relies on the fact that peculiar motions do not
affect tangential distances, but only the radial ones.
If we observe
the three-dimensional distribution of galaxies in redshift-space,
we can see that it is distorted with respect to that seen in real space:
dense clusters appear elongated along the line of sight ('fingers of God'
effect) and the correlation functions of galaxies are also distorted
when they are viewed in redshift-space. The important point is that
the redshift-space distortion effects can be used to get useful
information on important quantities like the PVD.
This can be done
using the anisotropies created by the peculiar
motions in the redshift space correlation function.
From a redshift survey of galaxies, it is possible to measure the correlation
function in redshift-space, $\xi(s)$. One can calculate $\xi(s)$ as function
of the component of the separation vector in the direction
parallel ($\pi$) and perpendicular ($r_{\rm p}$) to the line of sight.
Plotting the correlation function $\xi_{\rm z}(r_{\rm p},\pi)$
in the $\pi$-$r_{\rm p}$ plane,
redshift distortions give rise to:\\
1) a stretching of the contours of $\xi_{\rm z}(r_{\rm p},\pi)$
along $\pi$ on small scales
($<$ a few Mpc) because of non-linear pairwise velocities; \\
2) a compression along the line of sight, $\pi$,
on larger scales, because of bulk motions. \\
DP83 used the small scale distortion to get the
PVD of galaxies, modelling the elongation of $\xi_{\rm z}(r_{\rm p},\pi)$
along $\pi$ as a convolution of the real-space correlation function with
the distribution function $f(v_{12})$, being $v_{12}$ the relative
velocity of the galaxy pair in the direction of the line of sight.
The quoted method, introduced by Davis et al. (1978), Peebles (1980),
DP83  and Bean et al. (1983) has sometimes given 
discrepant values: DP83, using data from the CfA1,
obtained a value of $\sigma_{12}=340 \pm 40 {\rm km/s}$, at
$r=1 h^{-1} {\rm Mpc}$;
Fisher et al. (1994a) using the IRAS 1.2 Jy survey, obtained a value
of $\sigma_{12}=317^{+40}_{-49} {\rm km/s}$
in
agreement with that of DP83. Larger values was obtained by
Marzke et al. (1995) using the CfA2 and the SSRS2 (Southern Sky
Redshift Survey) redshift surveys,
($\sigma_{12}=540 \pm 180 {\rm km/s}$), by Lin et al. (1995)
($\sigma_{12}=452 \pm 60 {\rm km/s}$) using the Las Campanas Redshift Survey
and by J98 ($\sigma_{12}=570 \pm 80 {\rm km/s}$).
Moreover,
some authors (Mo et al. 1993; Marzke et al. 1995; Guzzo et al. 1997) have
pointed out that the value of the PVD is very sensitive to the presence
or absence of rich clusters in a sample. Other authors emphasized that
being $\sigma_{12}$ a pair-weighted statistic it is heavily weighted by the
densest regions of a sample (Strauss 1997). 
In two recent papers, Sheth et al. (2000) and Jing et al. (2001) have
shown that $\sigma_{12}$ varies widely not only
because of the previous effect, but also because of the morphological
type of the galaxies entering the catalog. In particular,
Sheth et al. (2000),
showed that redshift space distorsions should affect red
galaxies more strongly than blue.
This result can partly explain the difference
between the IRAS catalog $\sigma_{12}$ and the $\sigma_{12}$
computed from the more fair sample contained in the Las
Campanas Redshift Survey (LCRS).
However the most recent papers agree to values of the PVD in the range
400-600 km/s (J98).
This final range of convergence of $\sigma_{12}$, although
still large, comes because of the increasingly large volumes
surveyed in the most recent work.\\
\indent The $\sigma_{12}$ statistics has been used as a 
discriminator of
models. DP83's result compared with the SCDM model showed that 
the model predicts values of PVD larger than the observed ones and
successive studies confirmed that the SCDM model disagrees
with the observed values by a factor $3-6$ (Mo et al. 1993; Fisher
et al. 1994b; Marzke et al. 1995).
A new idea,
that of biased galaxy formation (Kaiser 1984;
Bardeen et al. 1986 (hereafter BBKS)),
was needed to reduce the discrepancy between
the SCDM model prediction and observations. Some recent studies have however
shown that for the SCDM model the shape of the predicted PVD is different
from that observed,
since the SCDM model predicts a
correlation function steeper than the observed one (J98).
In any case, introducing a constant bias does not resolve the problem, and
a scale-dependent bias is necessary to reduce the quoted discrepancy
(J98).\\
\indent Another tool that has long been used to determine $\sigma_{12}$
is the Cosmic Virial Theorem (CVT) (Peebles 1976) relying on the assumption
of hydrodynamic equilibrium of galaxies on small scales. Several authors
(Fisher et al. 1994b; BB96; SJ97)
have pointed out some limitations of the theorem. BB96 discussed how the mass distribution around a
typical pair of galaxies affects the CVT predictions, while
SJ97
improved the theorem taking into account the effect of the finite size
of galaxies. These last authors showed that the CVT in its initial formulation
overestimates the small-scale PVD.\\
\indent In this paper, I'll study the effects of a non-local bias,
induced by non-radial motions (see DG98, DG99, DG00;
D99),
on the PVD by means of both the CVT and the distortions in
$\xi_z(r_{\rm p}, \pi)$.\\ 
The plan of the paper is the following: in section ~2, I calculate the PVD
using the CVT in the original form introduced by Peebles (1976) and
the improved version of SJ97 and taking account of non-radial motions. The
theoretical results are 
compared with observational data. In section ~3,
I calculate the PVD by means of the distortions in the redshift-space
correlation function and compare the results with observations. Section ~4 is
devoted to conclusions.

\section{PVD and the CVT}

\subsection{The CVT and the finite size of galaxies}

The CVT introduced by Peebles (1976) is fundamentally an equation
of hydrostatic equilibrium obtained assuming statistical equilibrium
between galaxy pairs, on small scales. The formula describing this
theorem, and giving the relative (one-dimensional) peculiar velocity
dispersion  as function of their separation $r$,
can be obtained by means of the second BBGKY hierarchy equation.
It can be expressed as:
\begin{equation}
\langle v^2_{12} (r)\rangle
={6 G \rho_{\rm b} \over \xi_\rho(r)}
\int_r^\infty {d r' \over r'}
\int d {\bf z} \, { {\bf  r}' \cdot {\bf z} \over z^3}
\zeta_\rho (r', z, |{\bf  r}' - {\bf  z}|) ,
\label{eq:cvt1}
\end{equation}
where $\rho_{\rm b}$ is the mean density of the universe, $\xi_{\rho}$ and
$\zeta_{\rho}$ are the two and three-point correlation function of mass,
respectively.
Assuming the forms for the two and three-point correlation function
of galaxies given by
Groth \& Peebles (1977) and DP83, namely:
\begin{equation}
\xi_{\rm g}=\left(\frac{r_0}{r}\right)^\gamma
\end{equation}
and
\begin{equation} 
\zeta_{\rm g}(r_1,r_2,r_3)=Q_{\rm g} \left[
\xi_{\rm g}(r_1) \xi_{\rm g}(r_2)+\xi_{\rm g}(r_2)\xi_{\rm g}(r_3)+
\xi_{\rm g}(r_3)\xi_{\rm g}(r_1)
\right]
\end{equation}
where $Q_{\rm g}=0.7 \pm 0.21$ (see BB96),
$r_0=(5.4 \pm0.3) h^{-1} {\rm Mpc}$, $\gamma=1.77\pm0.04$,
and assuming that the two and three-point correlation functions of
mass, $\xi_{\rm \rho}$ and $\zeta_{\rm \rho}$ follow the same scaling and
satisfy the relations:
\begin{equation} 
\xi_{\rm g}(r)=b^2
\xi_{\rm \rho}(r)
\end{equation}
and 
\begin{equation} 
\zeta_{\rm g}(r_1,r_2,r_3)=Q_{\rm g} b^4
\zeta_{\rm \rho}(r_1,r_2,r_3)/Q_{\rm \rho}
\end{equation}
being $b$ the bias parameter,
the CVT gives
(Peebles 1976; Suto 1993):
\begin{eqnarray}
\langle v^2_{12} (r) \rangle^{1/2}
= 1460
\sqrt{\Omega_0 Q_\rho \over 1.3 b^2}
\sqrt{I_0(\gamma) \over 33.2} 5.4^{\frac {\gamma-1.8}2} \left({r_0
\over 5.4 h^{-1} {\rm Mpc}} \right)^{\frac\gamma 2} \left({r \over 1 h^{-1} {\rm Mpc}}
\right)^{\frac{2-\gamma}2} {\rm km/s} 
\label{eq:cvt1d}
\end{eqnarray}
where
$I_0(\gamma)$ is a function given in
Peebles (1980) (equation~(75.12)) and $\Omega_0$ is the density parameter.
\begin{figure}[ht]
\vspace{302pt}
\caption[]{The threshold $\delta_{\rm c}$  as a function of the mass M,
through the peak height $\nu$, 
taking account of non-radial motions with the model of this paper (dashed line)
is
compared with the result of Sheth et al. (1999) (solid line), obtained
using an ellipsoidal collapse model.}
\end{figure}
Numerical and analytical studies in non-linear gravitational clustering
seem to indicate that $Q_{\rm \rho}$ can be approximated as a constant
in the range $0.5-2$ (Suto 1996), which depends very weakly on the underlying
cosmological model. We shall use the value $0.6$, that gives
the same predictions of BB96 uncorrected CVT 
(see their Fig. 1), which shall be
used for comparisons with our paper (note that also a low
value of this parameter gives prediction for the pairwise-peculiar velocity
dispersions much larger than observations).

Equation~(\ref{eq:cvt1d}) was improved by SJ97 by taking account of the
fact that galaxies are not point-like, using perturbation
analysis and numerical integration. They changed the CVT incorporating a
non-zero core radius, $r_{\rm c}$, and softening the gravitational
force by means of a Plummer law with softening radius $r_{\rm s}$ obtaining
the final result:
\begin{eqnarray}
&& \langle v^2_{12} (r) \rangle^{1/2}
= 1460
\, \sqrt{\Omega_0 Q_\rho \over 1.3 b^2}~
5.4^{\frac {\gamma-1.8}2} \left(\frac{r_0}
{5.4 h^{-1} {\rm Mpc}} \right)^{\frac\gamma 2} \nonumber \\
&& \qquad \qquad \times
\sqrt{I(r_{\rm c}/r,r_{\rm s}/r;\gamma) \over 33.2}
\left({r \over 1 h^{-1} {\rm Mpc} }
\right)^{\frac{2-\gamma}2} {\rm km/s}
\label{eq:cvtpowerlaw}
\end{eqnarray}
where $I(r_{\rm c}/r,r_{\rm s}/r;\gamma)$ is given in SJ97 (equation~(10)). This
result simply tells that by replacing $I_0(\gamma)$ with
$I(r_{\rm c}/r,r_{\rm s}/r;\gamma)$ the finite size of the galaxies
is automatically
taken into account. Moreover they showed, using COBE normalized
CDM models, that the quoted improvement of the
CVT has as final result a reduction of the small-scale velocity dispersion
of galaxies, but not enough to be in agreement with observations.\\
\begin{figure}[ht]
\vspace{302pt}
\caption[]{The bias factor $b(\nu)$ as function of $\nu^2$.
The
solid line represents the spherical collapse prediction of Mo \& White (1996), the
dotted line the prediction for $b$ obtained from the model of this
paper and the dashed
line the ellipsoidal collapse prediction of Sheth et al. (1999). 
}
\end{figure}
\indent The final aim of this section is to determine the PVD, using the
modified CVT and, moreover, taking into account non-radial motions.

\subsection{The threshold $\delta_{\rm c}$, the selection function and bias}

The goal quoted at the end of the previous subsection 
can be accomplished remembering that non-radial motions induce a non-local
scale dependence of the bias parameter, $b$ (see D99))
and taking into account this effect in equation~(\ref{eq:cvtpowerlaw}). \\
In this subsection, we are going to show how to calculate the dependence
of $b$ on the peak height $\nu=\frac{\delta(0)}{\sigma}$, which is
proportional,
through the overdensity $\delta=\frac{\rho-\rho_{\rm b}}{\rho_{\rm b}}$ and
$\sigma$, the rms value of $\delta$, 
to the halo mass.
The $b(\nu)$ dependence obtained shall be
used in the next subsection to analyse the effects of non-radial motions
on the PVD.

\indent As shown by DG98, DG99, DG00,
if non-radial motions are
taken
into account, the threshold $\delta_{\rm c}$ is not a constant but is function of
mass, $M$, (DG98, DG99, DG00):
\begin{equation}
\delta _{\rm c}(\nu )=\delta _{\rm co}
\left[ 1+\frac{8G^2}{\Omega _o^3H_0^6r_{\rm i}^{10}%
\overline{\delta} (1+\overline{\delta} )^3}\int_{a_{\rm min}}^{a_{\rm max }}
\frac{L^2 \cdot da}{a^3}\right]
\label{eq:ma8}
\end{equation}
where $\delta _{\rm co}=1.68$ is the critical threshold for a spherical model, $%
r_{\rm i}$ is the initial radius, $L$ the angular momentum, $H_0$ and $\Omega_0$
the Hubble constant and the density parameter at the current epoch,
respectively,
$a$ the expansion parameter and $\overline{\delta}$ the mean fractional
density excess inside a shell of given radius and mass $M$.
The mass dependence of the
threshold parameter, $\delta_{\rm c}(\nu)$, and the total specific angular
momentum, $h(r,\nu )=L(r,\nu)/M$, acquired during expansion, were
obtained in the same way as described in D99 (see
also Fig. 1 of the quoted paper).\\
The result of the calculation is shown
in Fig. 1, where I plot $\delta_{\rm c}(\nu)$ obtained by
means of the model of this paper together with that obtained by Sheth et al. (1999) using
an ellipsoidal collapse model. The dashed line
represents $\delta_{\rm c}(\nu)$ obtained with the present model, while the
solid line that of Sheth et al. (1999). Both models show that the threshold
for collapse
decreases with mass.
In other words, this means that, in order to form
structure, more massive peaks must
cross a lower threshold,
$\delta_c(\nu)$, with respect to under-dense ones.
At the same time, since the
probability to find high peaks is larger in more dense regions, 
this means that, statistically, in order to form structure, 
peaks in more dense
regions may have a lower value of the threshold, $\delta_c(\nu)$, with respect
to those of under-dense regions.
This is due to
the fact that less massive objects are more influenced by external tides, and
consequently they must be more overdense to collapse by a given time.
In fact,
the angular momentum acquired by a shell centred on a peak
in the CDM density distribution is anti-correlated with density: high-density
peaks acquire less angular momentum than low-density peaks
(Hoffman 1986; Ryden 1988).
A larger amount of
angular momentum acquired by low-density peaks
(with respect to the high-density ones)
implies that these peaks can more easily resist
gravitational collapse and consequently it is more difficult for them to form
structure.
This is in agreement with  
Audit et al. (1997), Peebles (1990) and DP98, which pointed out that the
gravitational collapse 
is slowed down by the  effect of the shear
rather than fastened by it (as substained by other authors).
Therefore, on small scales, where the shear is statistically greater,
structures need, on average, a higher  density contrast to collapse.

This results in a tendency for less dense
regions to accrete less mass, with respect to a classical spherical model,
inducing a {\it biasing} of over-dense regions towards higher mass.
Moreover the scale dependence of the threshold, $\delta_{\rm c}(\nu)$,
implies also a scale dependence of the
bias parameter, $b$, since the two parameters are connected
(see Borgani 1990;
Mo \& White 1996; DG98, D99).
In fact, according to the biased theory of galaxy formation, observable
objects of mass $\simeq M$
arise from fluctuations of the density field, filtered on a scale
$R_{\rm f}$,
rising over a {\it global} threshold, $\delta>\delta_{\rm c}=\nu_t \sigma$,
where
$\sigma$ is the rms value of $\delta$ and $\nu_{\rm t}$ is the threshold
height.
The number density of objects, $n_{\rm pk}$,
that forms from peaks of density
of height $\nu$
can
be written, following BBKS, in the form:
\begin{equation}
n_{\rm pk}=\int_{0}^{\infty} t(\frac{\nu}{\nu_{\rm t}}) N_{pk}(\nu) d\nu
\end{equation}
where $t(\frac{\nu}{\nu_{\rm t}})$ is the threshold function,
and $N_{\rm pk}d\nu$ the differential number density of peaks
(see BBKS - equation~(4.3)). The threshold level $\nu_{\rm t}$ is defined so
that the probability of a peak becoming an observable object is $1/2$ when
$\nu=\nu_{\rm t}$. In the sharp threshold case the selection
function, is a Heaviside function
$t(\frac{\nu}{\nu_{\rm t}})=\theta(\nu-\nu_{\rm t})$.
The threshold function is connected to the bias
coefficient of a class of objects by (BBKS):
\begin{equation}
b(R_{\rm f})= \frac{\langle \tilde\nu \rangle }{\sigma_o}+1
\end{equation}
where $\langle \tilde\nu \rangle$ is:
\begin{equation}
\langle \tilde\nu \rangle  = \int_0^\infty
\left[ \nu -\frac{\gamma \theta }{%
1-\gamma ^2}\right] t(\frac{\nu}{\nu_{\rm t}}) N_{\rm pk}(\nu ) d\nu  \label{eq:nu}
\end{equation}
while, $\gamma $ and $\vartheta $ are given in BBKS
(respectively equation~(4.6a); equation~(6.14)).\\
The threshold (or selection) function can be obtained
following the arguments given in
Colafrancesco et al. (1995)
and  DG98.
In this last paper the selection function is defined as:
\begin{equation}
t(\nu )=\int_{\delta _c}^\infty p\left[ \overline{\delta} ,
\langle \overline{\delta} (r_{Mt},\nu
)\rangle ,\sigma _{\overline{\delta}} (r_{Mt},\nu )\right] d\delta
\label{eq:sel}
\end{equation}
where $\overline{\delta}$  is the mean fractional density excess inside
a given radius, as measured at the current epoch, assuming linear growth,
$<\overline{\delta}>$ its average value (see Ryden \& Gunn 1987) and
$\sigma_{\overline{\delta}}$ its dispersion given in Lilje \& Lahav (1991),
and
where the function
\begin{equation}
p\left[ \overline{\delta} ,\langle \overline{\delta} (r) \rangle\right] =
\frac 1{\sqrt{2\pi }\sigma_{\overline{\delta}} }\exp
\left( -\frac{|\overline{\delta} -
\langle \overline{\delta} (r) \rangle|^2}{2\sigma
_{\overline{\delta}} ^2}\right) \label{eq:gau}
\end{equation}
gives the probability that the peak overdensity, $\overline{\delta}$ is
different from the average, in a Gaussian density field.
The selection
function depends on $\nu $ through the dependence of $\overline{\delta} (r)$
on $\nu $.
As displayed, the integrand is evaluated at a radius $r_{\rm Mt}$ which is the
typical radius of the object that we are selecting. Moreover, the selection
function $t(\nu )$ depends on the critical overdensity threshold for the
collapse, $\delta _{\rm c}(\nu)$.
Given $\delta_{c}(\nu)$ and chosen a spectrum, the selection function is
immediately obtained through equation~(\ref{eq:sel}) and
equation~(\ref{eq:gau}).
As shown in DG98 (see their Fig. 6), the selection
function, as expected, differs from an Heaviside function (sharp threshold).
The shape of the selection function depends on the values of the
filtering length $R_{f}$ and on non-radial motions.
The value of $\nu$ at which the selection function $ t(\nu)$ reaches the
value 1 ($t(\nu) \simeq 1$) increases for growing values of the filtering
radius, $R_{\rm f}$.
This is due to the smoothing effect of the filtering process.
The effect of non-radial motions is, firstly, that of shifting
$t(\nu)$ towards higher values of $\nu$, and, secondly, that of
making it steeper.

This means that while in a $\theta$ threshold scheme, fluctuations below
$\delta_{\rm c}(\nu)$
have zero
probability to develop an observable object, and fluctuations above
$\delta_{\rm c}(\nu)$
have zero probability not to develop an object, the situation is totally
different when the threshold is not a sharp step function.
In this case objects can also be formed from fluctuations
below $\delta_{\rm c}(\nu)$ and
there is a non-zero probability for fluctuations above $\delta_{\rm c}$ to
be sterile. This result, in agreement with Borgani (1990), with the
'fuzzy' threshold approach of Audit et al. (1997) and Sheth et al. (1999),
is fundamentally connected
to non-sphericity effects present during the gravitational growing process.

I want to remark
that the choice made to calculate $b$, 
instead of the well known Mo \& White (1996) model:
\begin{equation}
b=1+\frac{\nu^2-1}{\delta_{\rm c}}
\label{eq:bias}
\end{equation}
is motivated by the fact that I am interested in studying
the PVD on scales $<10  h^{-1} {\rm Mpc}$,
while in their model, Mo \& White (1996) obtained analytic
results only in the limit of large separations (Sheth \& Lemson 1999,
Taruya \& Suto 2000).
Although not directly necessary to the development of the remaining
part of the paper, I also calculate the large scale bias factor, $b$,
in order to test the model described and to compare it with the value found by Sheth et al. (1999), as
previously done with the threshold, $\delta_{\rm c}$.

In Fig. 2, I plot the bias parameter, $b$, as a
function of the peak height
$\nu$, which is proportional to the halo mass. The
solid line shows the spherical collapse prediction of Mo \& White (1996), the
dotted line the prediction for $b$ obtained from our model, and the dashed
line the ellipsoidal collapse prediction of Sheth et al. (1999). As shown in
the figure, the effect of non-radial motions is to change
the dependence
of $b$ on $\nu$ in good agreement with Sheth et al. (1999).
From Fig. 2 it is evident that at the low mass end the bias relation has an
upturn, meaning that less massive haloes are more strongly clustered
than the prediction of the spherical collapse model 
of Mo \& White (1996) and in agreement with N-body simulations (Jing 1998;
Sheth \& Lemson 1999).

\subsection{PVD, CVT and non-radial motions}

At this point I am ready to calculate the PVD, using the CVT in its original
formulation (Peebles 1976) and the improved version of SJ97, and to study
the effect of non-radial motions on these results.

Fig. 3 plots the results of the model of this paper and the comparison with
the DP83 data for values of
the parameter 
$F=0.1, 1, 1.5$ (open squares), the IRAS value at $1 h^{-1} {\rm Mpc}$
(filled square)
(Fisher et al. 1993), the DP83 interpretation of the Turner (1976)
galaxy pair catalog (solid hexagons) and the Marzke et al. (1995) data at
$1 h^{-1} {\rm Mpc}$
(dashed errorbar).
\footnote{We recall that the parameter $F$ was introduced in the infall
model used by DP83 (see DP83 equation~(22)-(23))
in order to test the sensitivity
of the velocity dispersion estimates to the streaming correction.
According to the equation~(23) in DP83,
a small value of $F$ implies that galaxies on all scales
expand with the Hubble flow, while a value of $F=1$, implies, 
that galaxies having separation of
$5 {\rm Mpc}$ are expanding at $1/2$ the Hubble rate.}
Here, the PVD, is derived using
a SCDM model ($\Omega_0=1$, $h=1/2$) filtered on galactic scales and
normalized to
$Q_{\rm COBE} = 17 {\rm \mu K}$.
The dotted line represents the PVD obtained from the uncorrected version of
the CVT. 

As previously found by several authors (DP83; Mo et al. 1993;
Fisher et al. 1994b; Marzke et al. 1995), the SCDM model overestimates
the PVD. The dashed line represents the PVD obtained from the corrected
version of the CVT (SJ97), having assumed
$r_{\rm c}=r_{\rm s}=10 h^{-1}{\rm kpc}$.

In agreement with SJ97,
taking into account the finite size of galaxies reduces the small-scale velocity
dispersion of galaxies, but not enough to be in agreement with observational
data. The solid line represents
the PVD, taking account of the non-radial motions, while the
long-dashed line takes also into account 
the finite size of galaxies (again $r_{\rm c}=r_{\rm s}=10 h^{-1}$ Kpc).
\begin{figure}[ht]
\vspace{302pt}
\caption[]{The PVD as a function of separation $r$.
The PVD, is derived using
a SCDM model ($\Omega_0=1$, $h=1/2$), whose transfer
function is given in next section, equation~(\ref{eq:ma5}),
filtered on galactic scales and
normalized to $Q_{\rm COBE} = 17 \mu$ K.
The dotted line represents the PVD obtained from the uncorrected version of
the CVT, the dashed line represents the PVD obtained from the corrected
version of the CVT (SJ97) with $r_{\rm c}=r_{\rm s}=10 h^{-1}{\rm kpc}$,
the solid line
the PVD taking into account the non-radial motions while the
long-dashed line takes also into account 
the finite size of galaxies ($r_{\rm c}=r_{\rm s}=10 h^{-1}{\rm kpc}$).
The model is compared with DP83's data for values of
the parameter
$F=0.1, 1, 1.5$ (open squares), the IRAS value at $1 h^{-1} {\rm Mpc}$
(filled square)
(Fisher et al. 1993), the DP83 interpretation of the Turner (1976)
galaxy pair catalog (solid hexagons) and the Marzke et al. (1995) data
(dashed errorbar).
}
\end{figure}
The finite size effect and non-radial motions both produce a decrease in the
PVD. This is due to the fact that
the finite size effect suppresses the effective gravitational force
between pairs and changes the two-point correlation function
on small scales.
Non-radial
motions have also the effect of changing the
two-point correlation function (DG99; D99),
the mass distribution and the density
profile of the galactic halos (White \& Zaritsky 1992;
DG98). The change of mass distribution
produces, in agreement with BB96, a change
in the PVD predicted by the CVT,  
namely
smaller values of the small-scale pairwise peculiar velocity dispersions
of galaxies. Differently from BB96 in the model of this paper,
I do not observe the rapid
increase of the PVD beyond $r \simeq 1 h^{-1} {\rm Mpc}$ with a consequent
better agreement of the model with data.
The difference is due to the increase of the bias parameter going from
small to larger scales. Aside from the model introduced to calculate $b$,
the behavior of the bias parameter, and then
the PVD, can be qualitatively explained as follows:
the papers of Sheth (1996) and Diaferio \& Geller (1996) and Mo et al. (1997) 
suggest that, while on very small separations, pairs come
from both small and large haloes, at larger separations, pairs come
mainly from larger and more massive haloes.  Since massive haloes have
larger $b$ values than less massive ones, this means that the
effective bias is lower on small scales than on larger ones.

The CVT has been traditionally applied as an indicator of the
cosmological density parameter, $\Omega_0$, and since
its first use (DP83), it was considered as a strong indicator
of an open universe. Excluding the behaviour of the PVD at
$r \ge 1 h^{-1} {\rm Mpc}$, where,
as remarked by BB96, the stable clustering hypothesis could be questioned, 
the result of this
paper is in agreement with that of BB96, namely:
the PVD should not be taken
as a direct evidence of a low density universe. At the light
of Fig. 3, I want to add
that this conclusion should be more strict if we could rely on the
CVT for $r> 1 h^{-1} {\rm Mpc}$.
At the same time, as
remarked by SJ97, the usefulness of the CVT to put constraint on $\Omega_0$
is strictly connected with the quality of the observational data. 
In fact, as reported in the introduction, 
some authors (Mo et al. 1993; Marzke et al. 1995; Guzzo et al. 1997) have
pointed out that the value of the PVD is very sensitive to the presence
or absence of rich clusters in a sample.
There is a large variation in $<v^2_{12}>$ in different
surveys and even for the same survey analyzed in different ways.
Moreover, $\sigma_{12}$ varies widely
also because of the morphological
type of the galaxies entering the catalog (Sheth et al. 2000).
As shown by Mo et al. (1997), the numbers of pairs with small separations is
dominated by haloes having $M \simeq M_{\ast}$ while on small scales massive
pairs, $M > M_{\ast}$, strongly affect the PVD. Since the number density, $n$,
of this massive haloes is small, a large sample is needed to have a fair
sample of $n$ or in other terms the small scale PVD can be fairly sampled
only by means of samples containing many rich clusters.
%
%

\section{PVD from distortions in $\xi_z(r_{\rm p}, \pi)$}

Since the combination of statistical and systematic uncertainties
entering the CVT have lead several authors to conclude that it 
has notheworthy problems 
as an estimator of cosmological parameters
(Fisher et al. 1994b), in order to be on the safe side, I also use a more reliable and more widely
used method to
estimate cosmological parameters, and in particular the PVD. 

As previously reported in the introduction, the three-dimensional distribution
of galaxies in redshift-space appears distorted with respect to the same
distribution in real space. The correlation function measured in
redshift space, $\xi(s)$, is different from the real space counterpart,
$\xi(r)$, because of two effects.
These effects are seen in redshift-space, plotting the correlation function
in terms of two variables, the separations parallel ($\pi$) and
perpendicular ($r_{\rm p}$) to the line of sight. Given a pair of galaxies
with redshifts corresponding to velocities ${\bf v}_1$ and ${\bf v}_2$, the
separation in redshift space is given by:
\begin{equation}
{\bf s}={\bf v}_1-{\bf v}_2
\label{eq:pri}
\end{equation}
while the observer's line of sight is:
\begin{equation}
{\bf l}=\frac{1}{2} ({\bf v}_1+{\bf v}_2)
\end{equation}
The separations parallel and perpendicular to the line of sight are
respectively:
\begin{equation}
{\bf \pi}=\frac{{\bf s} {\bf l}}{|\bf l|} 
\end{equation}
and 
\begin{equation}
{\bf r}^2_{\rm p}={\bf s} {\bf s} - \pi^2 
\label{eq:ult}
\end{equation}
The PVD can be obtained by modelling the redshift distortion of
$\xi_z(r_{\rm p}, \pi)$, 
as follows:

The statistics $\xi_z(r_{\rm p}, \pi)$ is a convolution of the real-space
correlation function, $\xi(r)$, with the distribution function of
the relative velocity along the line of sight, $f(v_{12})$.
Following, for example, 
Fisher et al. 1994b, 
Jing \& B\"orner (1998) (hereafter JB98), we have:
\begin{equation}
1+\xi_z(r_{\rm p}, \pi)=\int f(v_{12}) \left[1+
\xi\left(\sqrt{r^2_{\rm p}+(\pi-v_{12}/H_0)^2}\right)\right] d v_{12}
\label{eq:xizxir}
\end{equation}                                

To obtain the PVD from $\xi_z(r_{\rm p}, \pi)$
there are at least three methods (Peebles 1980, section~76).
The first one, which I am going to use in this paper and which
is the more diffused in papers that try to recover
the PVD from survey of galaxy redshift (Davis \& Peebles 1983; Mo et al. 1993;
J98; JB98), can be summarized in the following steps:\\
1) Estimate of the redshift-space two-point correlation function
$\xi_z(r_{\rm p}, \pi)$. If one has observational data, coming from a redshift
survey, it is necessary to choose an estimator for $\xi_z(r_{\rm p}, \pi)$ (see
for example J98). In our case, $\xi_z(r_{\rm p}, \pi)$ is to be calculated
theoretically from the power spectrum, as I am going to do.\\
2) Estimate of the projected two-point correlation function $w(r_{\rm p})$:
\begin{equation}
w(r_{\rm p})=2 \int^\infty_0 \xi_{\rm z} (r_{\rm p},\pi) d \pi=
2 \int^\infty_0 \xi(\sqrt{r_{\rm p}^2+y^2})d y 
\end{equation}
3) Calculation of the two-point correlation function $\xi(r)$.\\
4) Assumption of a functional form for $f(v_{12})$, an infall model for
$\overline{v}_{12}$ to, finally, solve equation (\ref{eq:xizxir}) for
$\sigma_{12}$.\\

For seek of completeness, I shortly describe the other two methods. 
The second approach follows from Eq. \ref{eq:xizxir}. After determining
$\xi(r)$ from:
\begin{equation}
\int^\infty_0 \xi_{\rm z} (r_{\rm p},\pi)=
\int^\infty_0 \xi(\sqrt{r_{\rm p}^2+y^2})d y
\end{equation}
the PVD can be obtained by the equation:
\begin{equation}
<v^2_{12}>
= 3 H_0^2 \int^\infty_0 d \pi \pi^2 \left[
\xi_{\rm z} (r_{\rm p},\pi)-\xi(\sqrt{r_{\rm p}^2+y^2})
\right]/
\int^\infty_0 \xi_{\rm z} (r_{\rm p},\pi) d \pi
\end{equation}
(see also Peebles 1979).
This method does not require the knowledge of
$f(v_{12})$ but, as noted by Davis \& Peebles (1983), the integral depends
sensitively
on $\xi$ at $|\pi| \geq 1000 {\rm km/s}$, where, if one uses data from
redshift surveys,
$\xi_{\rm z} (r_{\rm p},\pi)$ is poorly known.
The third method also deriving from Eq. \ref{eq:xizxir} is:
\begin{equation}
<v^2_{12}>
= 3 H_0^2 \int^\infty_0 d \pi d r_{\rm p} (\pi^2-r_{\rm p})
\xi_{\rm z} (r_{\rm p},\pi)/
\int^\infty_0 \xi_{\rm z} (r_{\rm p},\pi) d \pi d r_{\rm p}
\end{equation}
which requires only $\xi_{\rm z} (r_{\rm p},\pi)$ but also this approach, as
the previous one, is not free from drawbacks (see Peebles 1980, Sect. 76). \\
Coming back to the first method, we can start from the step (1).

In order to get
$\xi_{\rm z}(r_{\rm p}, \pi)$, I Fourier transform the redshift-space
power spectrum:
\begin{equation}
\xi(s,\mu_{{\bf s l}})^S=\frac{1}{(2 \pi)^3} \int d^3 {\bf k}
P^S(k, \mu_{{\bf k l}}) \exp{(-i k s \mu_{{\bf k s}})}
\label{eq:xiss}
\end{equation}
(see Cole et al. 1994, equation~(B1)), 
where, as previously quoted, ${\bf l}$ denotes the direction of the
line of sight, $\mu_{{\bf k s}}$ and $\mu_{{\bf k l}}=
\frac{k_{\parallel}}{(k_{\parallel}^2+k_{\perp})^{1/2}}$
the cosine between the vectors
${\bf s}$, ${\bf k}$ and ${\bf k}$, ${\bf l}$, respectively ($k_{\perp}$ and
$k_{\parallel}$ are the components of the wavevector perpendicular and
parallel to the line of sight).
$\xi_{\rm z}(r_{\rm p}, \pi)$ can be obtained from equation~(\ref{eq:xiss}) by
changing variables according to equation~(\ref{eq:pri})-(\ref{eq:ult}).

Before going on, I want to recall that there are three effects that cause
a departure of 
the observed clustering properties of galaxies from the linear spectrum:\\
a) Redshift space effects. We extract three dimensional
clustering information from redshift surveys; in these surveys,
the galaxy radial coordinates are distorted by peculiar
velocities.\\
b) Nonlinear evolution, producing, on small scales, a departure of the mass
spectrum from its initial form.\\
c) Bias. Different classes of objects trace mass in different ways leading
to difficulties in connecting theory and observations.

Firstly, (relatively to point a), in order to calculate
the redshift space power spectrum, distorted by the peculiar velocity field, I
use the expression given by Peacock \& Dodds (1994) and Cole et al. (1995):
\begin{equation}
P^S(k_{\perp}, k_{\parallel})=P^R(k) \left[1+\beta \mu^2_{\bf k l} \right]^2
\left(1+k^2 \sigma_{12}^2 \mu^2_{\bf k l}/2\right)^{-2}
\label{eq:redpe}
\end{equation}
where $k=\sqrt{k^2_{\parallel}+k^2_{\perp}}$, $P^R(k)$ is the real-space power
spectrum, and $\beta=\frac{f(\Omega)}{b}$, where the perturbation parameter
$f(\Omega)$ is defined in Peebles 1980 (section~14).
The real-space
power spectrum that I adopt is $P(k)=Ak T^2(k)$ 
with the transfer function $T(k)$ given in BBKS (equation~(G3)):
\begin{equation}
T(k) = \frac{[\ln \left( 1+2.34 q\right)]}{2.34 q}
\cdot [1+3.89q+ 
(16.1 q)^2+(5.46 q)^3+(6.71)^4]^{-1/4}
%
%
\label{eq:ma5}
\end{equation}
where $ A$ is the normalizing constant and $q=\frac{k
\theta^{1/2}}{\Omega_{\rm X} h^2 {\rm Mpc^{-1}}}$. 
Here $\theta=\rho_{\rm er}/(1.68 \rho_{\rm \gamma})$
represents the ratio of the energy density in relativistic particles to
that in photons ($\theta=1$ corresponds to photons and three flavors of
relativistic neutrinos).
The power spectrum was COBE normalized
using the cosmic microwave anisotropy quadrupole
$Q_{rms-PS}=17 {\rm \mu K} $ 
that corresponds to $\sigma_8=0.95 \pm 0.2$
(Smoot et al. 1992; Liddle \& Lyth 1993).  

Secondly, (relatively to point b),
since I have to calculate the PVD in the range
$0.01 \le r_{\rm p} \le 10 h^{-1} {\rm Mpc}$ I should take into account the
non-linear evolution of the power spectrum.

To this aim, I shall use the fitting formula of Peacock \& Dodds (1996)
(see also Hamilton et al. (1991), Peacock \& Dodds (1994), Jain et al. (1995),
Mo et al. (1997)),
relating
the evolved
power variance, $\Delta^2_{\rm E}= \frac{1}{2 \pi^2} k^3 P(k)_{E}$, to
the initial density spectrum:
\begin{equation}
\Delta^2_{\rm E}(k_{\rm E})= F([\Delta^2(k_{\rm L})])
\end{equation}
where L stands for 'linear' E for 'evolved',
$k_{\rm L}=[1+\Delta^2_{\rm E}(k_{\rm E})]^{-1/3} k_{\rm E}$ and
the functional form of $F$ is given in Peacock \& Dodds (1996).
This correction was done
to the real-space spectrum in equation~(\ref{eq:redpe}) which shall be introduced
in equation~(\ref{eq:xiss}) to obtain $\xi_{\rm z}(r_{\rm p},\pi)$.\\

After $\xi_z(r_{\rm p}, \pi)$ has been calculated, we can go to step
(2) and (3).

Following the usual technique used to recover
the real-space correlation function from $\xi_z(r_{\rm p}, \pi)$,
I define a projected
function $w(r_{\rm p})$, unaffected by redshift distortions:
\begin{equation}
w(r_{\rm p})=2 \int^\infty_0 \xi_{\rm z} (r_{\rm p},\pi) d \pi=
2 \int^\infty_0 \xi(\sqrt{r_{\rm p}^2+y^2})d y 
\end{equation}
(DP83).  Then $\xi(r)$ can be obtained, following
DP83, either solving $w(r_{\rm p})$ for $\xi(r)$, or also fitting 
$w(r_{\rm p})$ to a power law model for $\xi(r)$. Following the first
alternative $\xi(r)$ is given by evaluating the
Abel integral:
\begin{equation}
\xi(r)=-\frac{1}{\pi} \int_r^{\infty} dr_{\rm p} w(r_{\rm p})
(r^2_{\rm p}-r^2)^{-1/2}
\end{equation}
(DP83)
(an alternative method to calculate $\xi(r)$, even if more complex,
and compatible with that
previously presented, as remarked by Peebles (1983) and J98, should be that
presented in D99, and based on the use of the Limber's
equation. As shown in that case, the effect of non-radial motions is that
of producing enough additional clustering to fit the $\xi(r)$ of the APM
galaxy survey).

Finally, (step 4), I need a functional form for $f(v_{12})$.
A good approximation to this
function is:
\begin{equation}
f(v_{12})=\frac{1}{\sqrt{2} \sigma_{12}} \exp{(-\frac{\sqrt{2} |v_{12}
-\overline{v}_{12}|}{\sigma_{12}})}
\end{equation}
(Fisher et al. 1994b; J98), where
$\overline{v}_{12}$ and $\sigma_{12}$ are the mean and the
dispersion of the 1-D pairwise peculiar velocities.
\begin{figure}[ht]
\vspace{302pt}
\caption[]{The PVD calculated from the redshift-space distortion of
$\xi_{\rm z}(r_{\rm p},\pi)$.
The solid line is the PVD taking account of non-radial motions,
the dashed line the PVD obtained from the SCDM model that does not take
account of the non-radial motions.
The data are the PVD measured from the Las Campanas Redshift Survey by J98.
}
\end{figure}
The infall, $\overline{v}_{12}$, is difficult to model because it is correlated with
$\sigma_{12}$, and it is scale dependent.
Following J98,
I assume a self-similar infall model for $\overline{v}_{12}$:
\begin{equation}
\overline{v}_{12}({\bf r})=\frac{-y H_0}{1+(r/r_{\ast})^2}
\end{equation}
where $r_{\ast}=5 h^{-1} {\rm Mpc}$ and $y$ is the radial separation
in real space. This infall model is usually assumed, because it gives
a good approximation to the infall pattern seen in CDM models and, as
shown, in JB98, J98 the PVD reconstructed using this model
is in good agreement with the 3-D velocities in the simulations.

As stressed by J98, the method described to determine the
PVD is approximated for several reasons:   \\
1) the functional
forms of $f(v_{12})$ and $v_{12}(r)$ are approximations; \\
2) the
reconstruction of the PVD from redshift distortion gives, as a result, only an
average of the real PVD along the line of sight. \\
In any case, the value of $\sigma_{12}$,
obtained with the quoted technique, is within $20 \%$ of the true PVD
(JB98).

At this stage, I have all the quantities required by equation~(\ref{eq:xizxir}) to obtain
$\xi_z(r_{\rm p}, \pi)$.
I have obtained the PVD by using
equation~(\ref{eq:xizxir}) and 
fitting $\sigma_{12}$
to $\xi_{\rm z}(r_{\rm p},\pi)$
(see Peebles 1980; DP83; Mo et al. 1993; J98). The calculation was repeated
two times, the first with $\delta_{\rm c}=1.686$ and the second assuming
equation~(\ref{eq:ma8}).\\
The result of the model described in this section is shown in Fig. 4. Here
the solid line represents the PVD, taking into account non-radial motions,
the dashed line the PVD obtained from the SCDM model that does not take
into account the non-radial motions. The theoretical results are compared with
the PVD measured from the Las Campanas Redshift Survey by J98.
As shown in Fig. 4, the PVD predicted by the SCDM model that does not take
into account 
non-radial motions is higher than the observed value for all $r_{\rm p}$
except at $r_{\rm p}=30 h^{-1} {\rm Mpc}$. However, as pointed out by
J98, the results on scales larger than $5 h^{-1} {\rm Mpc}$
are very sensitive to the model for the infall chosen, since the
statistical fluctuations become very large.
The result confirms the well known limit of the CDM model to reproduce
the two-point correlation function, to which the PVD is connected, since
it probes the clustering power on small scales. As stressed several times,
the discrepancy of the $\xi(r)$ on scales $r_{\rm p} \geq 5 h^{-1} {\rm Mpc}$
is due to the fact that this model does not have enough power on large scales.
Unless the PVD of galaxies is biased relative to that of the mass, the CDM
model have problems in fitting the observed data.
Moreover, using
a constant bias, it is possible to fit the data only on a limited radius
range. Fig. 4 shows that the introduction of a scale-dependent bias, due to non-radial
motions, produces a notheworthy reduction of the discrepancy of the PVD
measured from the Las Campanas Redshift Survey with respect to the prediction
of the CDM model.
The mechanism giving rise to this bias was described in the previous section
and more widely in DG98, DG99, DG00, D99.
Before concluding, it is interesting to note that, J98, showed
that a scale-dependent bias may explain the discrepancy between the model
predictions and observational results. In agreement with J98,
a velocity bias is not needed to make models compatible with the observed PVD.

\section{Conclusions}

In this paper, I have studied the effect of non-radial motions on the PVD
in a SCDM model extending the model introduced in
DG98 and D99. Firstly, I calculated the effect of non-radial motions
on the threshold $\delta_{\rm c}$ and bias.
I then compared the large scale
bias, calculated by means of the model of this paper, with the prediction
of Mo \& White (1996) model for a spherical collapse model and with the
result of 
Sheth et al. (1999) for an ellipsoidal collapse, finding a good agreement
with the result of this last paper.
The model for the bias was used in the SJ97
improved version of CVT to determine the PVD, which was compared with 
the data from DP83,
the IRAS value at $1 h^{-1} {\rm Mpc}$ of
Fisher et al. (1993) and Marzke et al. (1995). 
As shown in Fig. 3, non-radial motions produce
a reduction of the values of the PVD, with respect to the prediction of
a SCDM model that does not take into account this effect. The result
is due to the scale-dependent non-local bias induced by non-radial motions
and to the change of mass distribution in galactic halos.
At the light
of BB96, the result may be also read as a strong dependence of the CVT
prediction on the model chosen to describe the mass distribution
around galaxies, suggesting 
that the CVT cannot be taken as a direct evidence of a low density
universe.

Finally, I calculated the PVD by means of the redshift distortion in the
redshift-space correlation function, $\xi_{\rm z}(r_{\rm p},\pi)$, and
compared it with the data obtained by J98 from the
Las Campanas Redshift Survey. The result confirmed that non-radial motions
reduce the discrepancy between SCDM model predictions and observations.

\section*{Acknowledgements}
I would like to thank Y.P. Jing, R. Sheth, A. Diaferio and
E. Nihal Ercan for some useful comments that helped me in improving this paper.
Finally, I would like to thank 
Bo$\breve{g}azi$\c{c}i University
Research Foundation for the financial support through the project code
01B304.


\newpage
\begin{figure}
\psfig{file=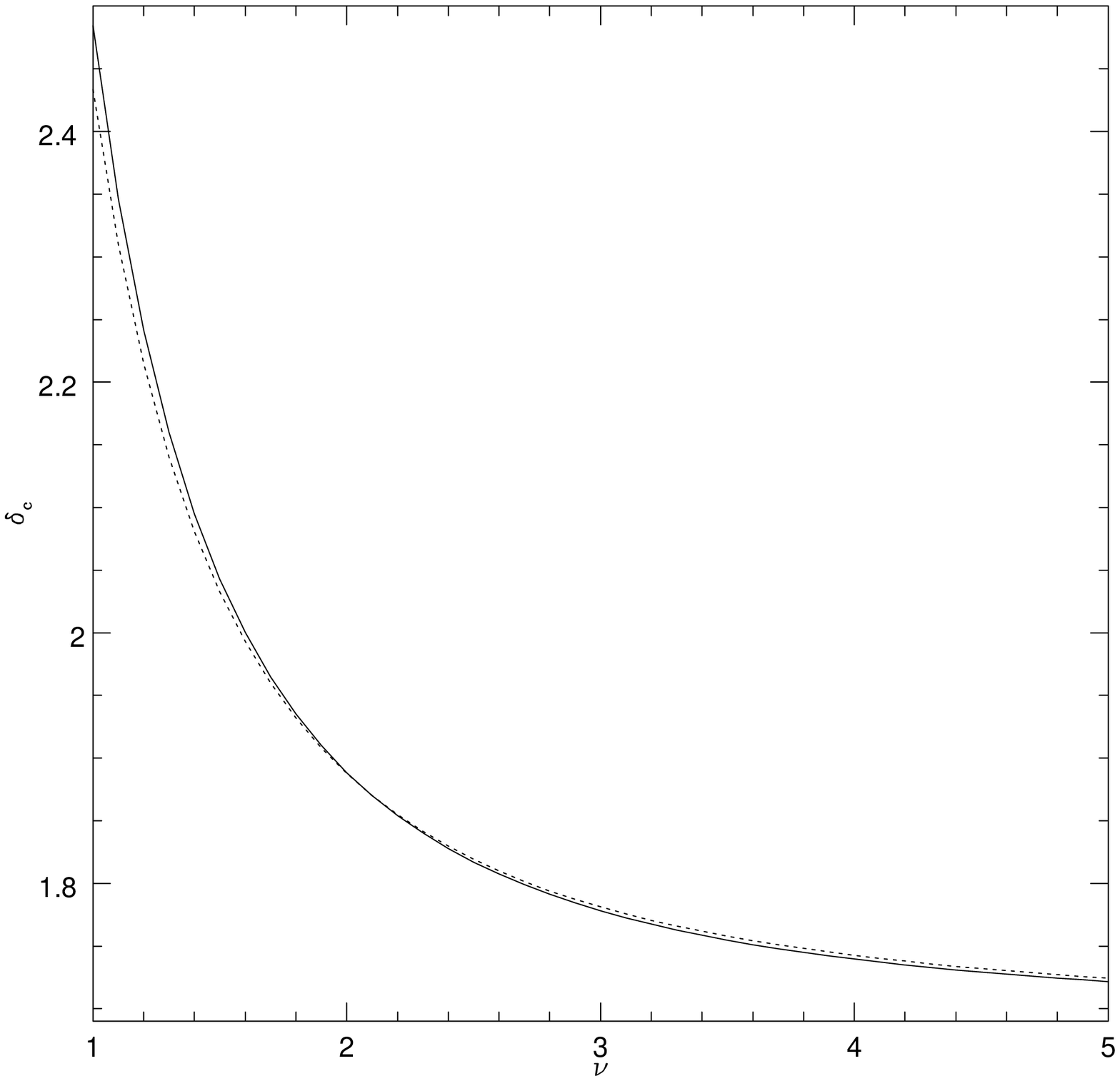,width=19cm}
\label{Fig. 1}
\end{figure}
\newpage
\begin{figure}
\psfig{file=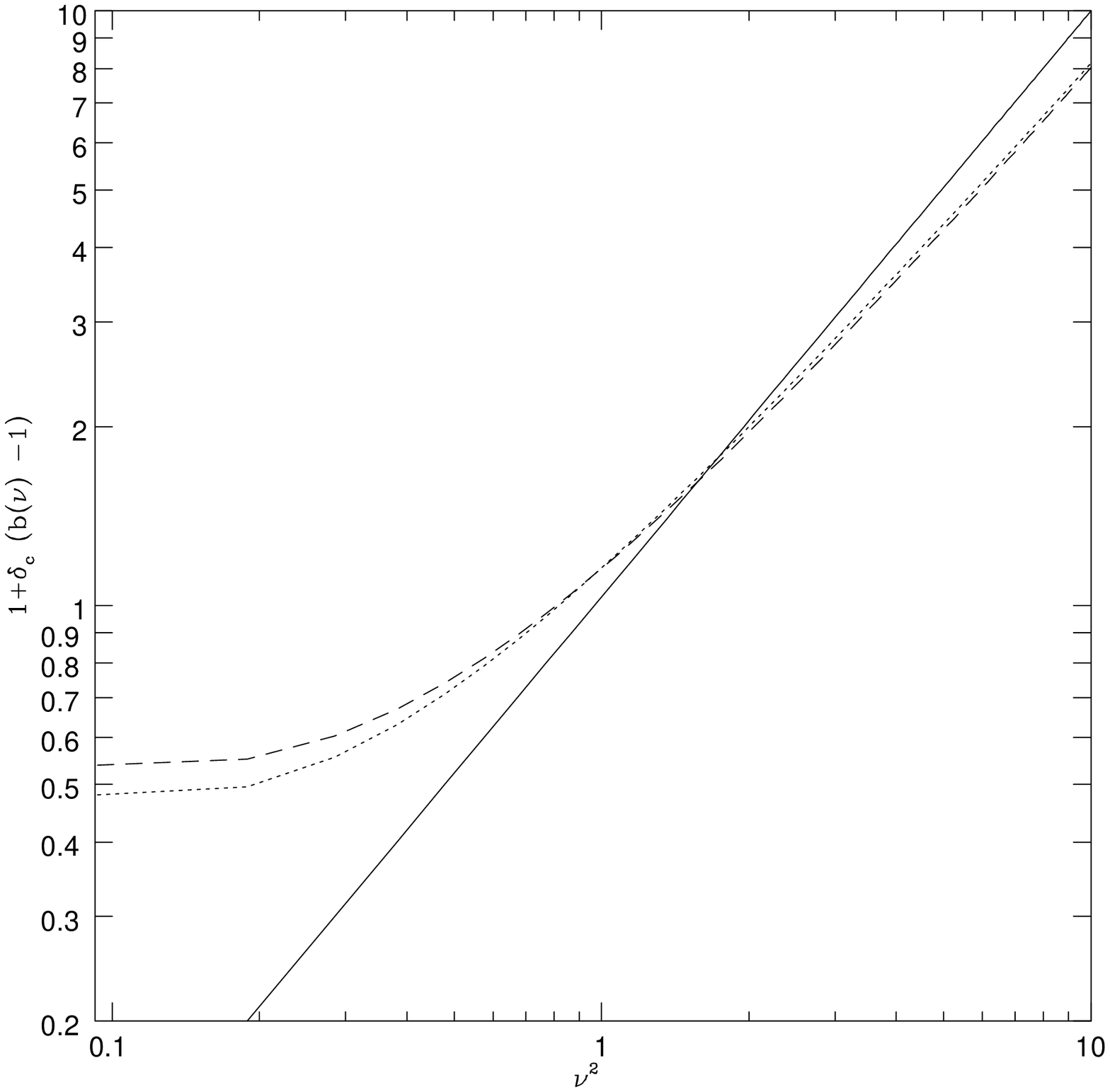,width=19cm}
\label{Fig. 2}
\end{figure}
\newpage
\newpage
\begin{figure}
\psfig{file=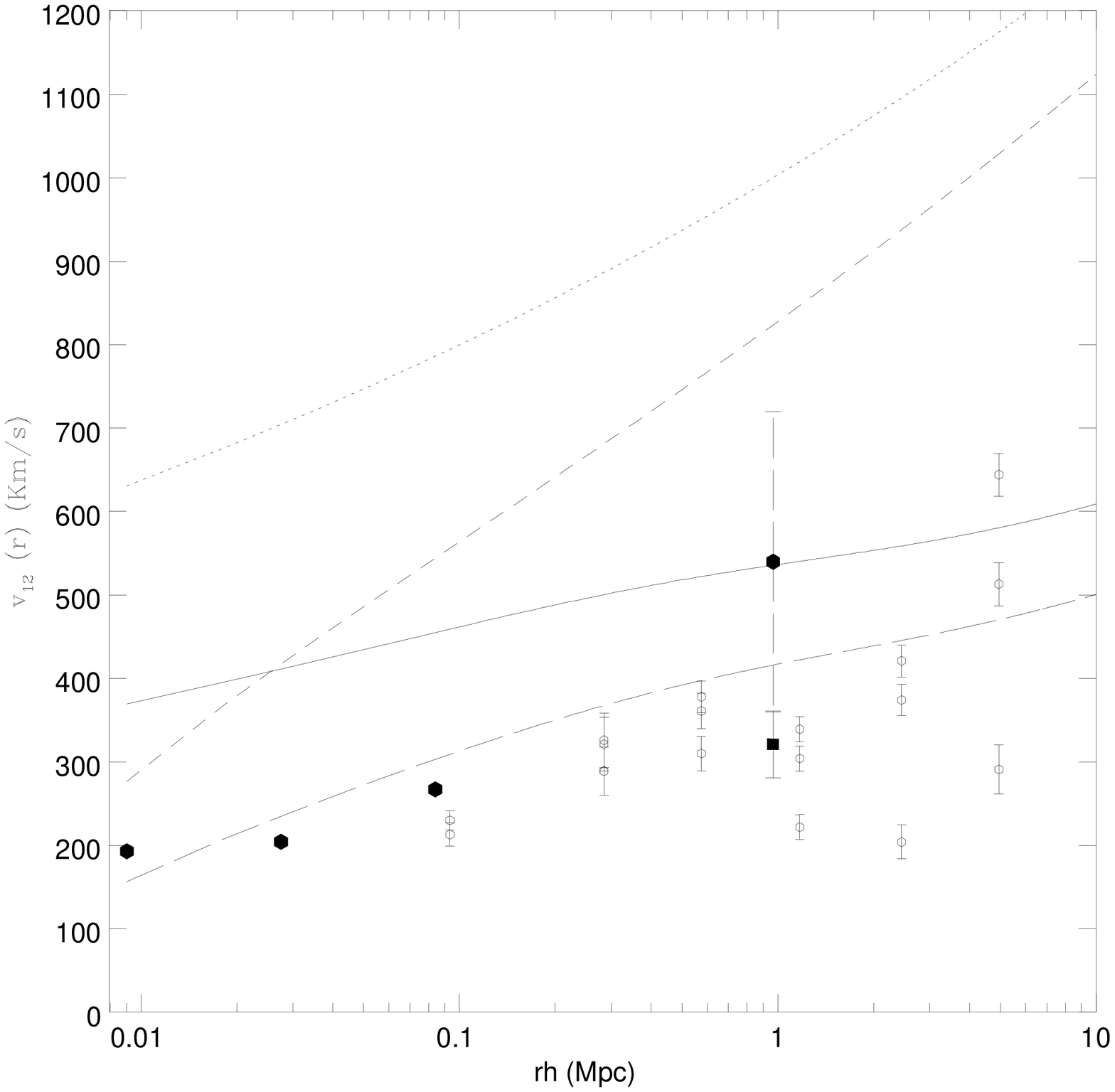,width=19cm}
\label{Fig. 3}
\end{figure}
\newpage
\begin{figure}
\psfig{file=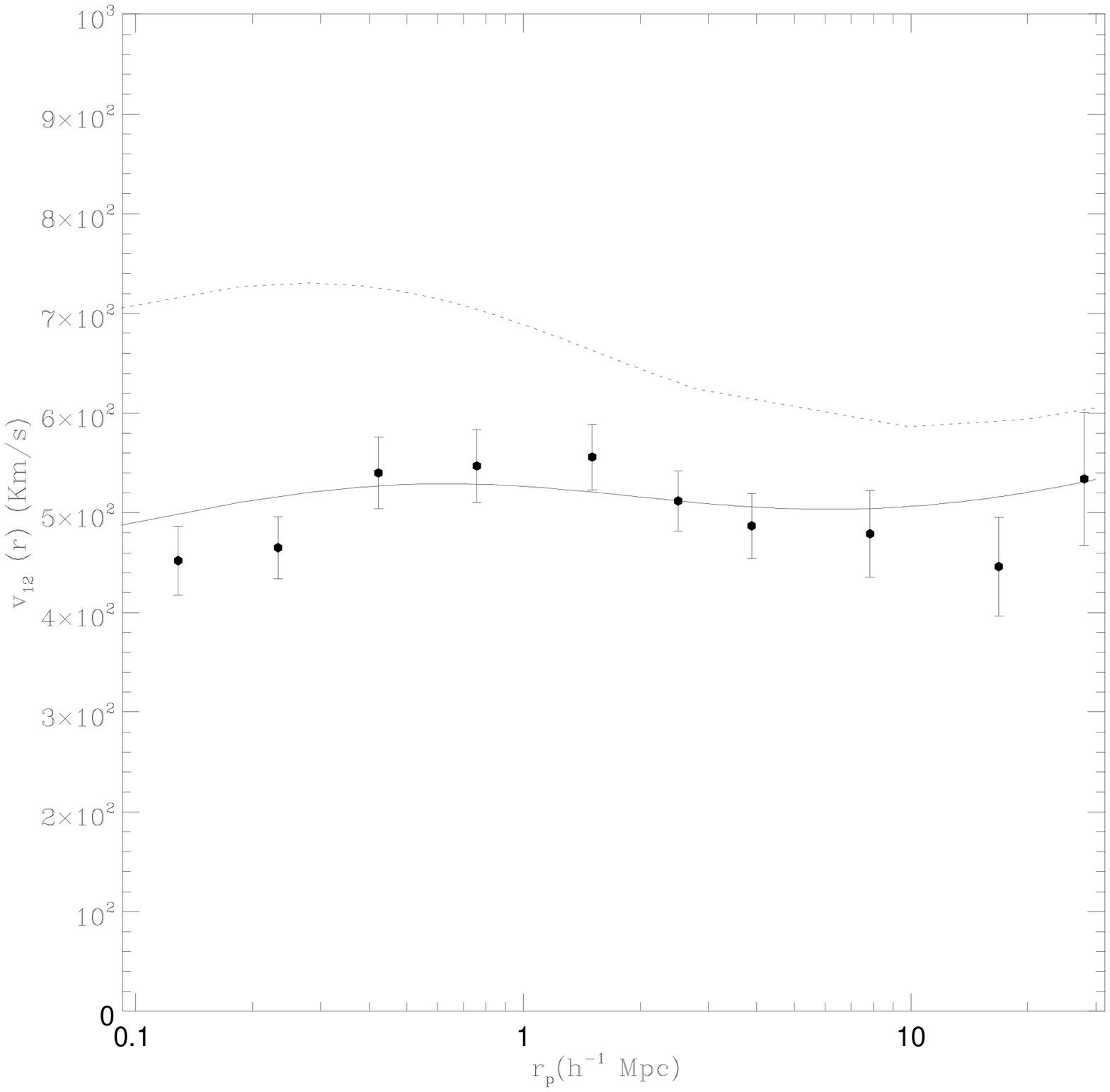,width=19cm}
\label{Fig. 4}
\end{figure}
\end{document}